\documentclass[figures]{epl}

\usepackage{rotating}
\usepackage{amsmath}
\usepackage{amssymb}
\usepackage{mathrsfs}
\usepackage{graphics}
\usepackage{graphicx}
\usepackage{epsf}

\title{Diffusion mechanisms of localised knots along a polymer}
\shorttitle{Knot diffusion}

\author{Ralf Metzler\inst{1}\inst{2}\thanks{E-mail: \email{metz@nordita.dk}},
Walter Reisner\inst{3}, Robert Riehn\inst{4}, Robert Austin\inst{4}, Jo\-nas
Tegenfeldt\inst{5}, \and Igor M. So\-ko\-lov\inst{6}}
\shortauthor{R.~Metzler \etal}
\institute{\inst{1} NORDITA -
                    Blegdamsvej 17, DK-2100 Copenhagen\\
           \inst{2} New address: Physics Dept,
                    University of Ottawa - Ottawa, ON, K1N 6N5, Canada\\
           \inst{3} Biophysics Dept, Ris{\o} National Laboratory,
                    Frederiksborgvej 399, DK-4000 Roskilde\\
           \inst{4} Department of Physics, Princeton University,
                        Princeton, NJ 08544\\
           \inst{5} Division of Solid State Physics, Lund University,
                    S{\"o}lvegatan 14, S-223 62 Lund\\
           \inst{6} Inst.~f{\"u}r Physik, Humboldt Universit{\"a}t zu
                    Berlin - Newtonstra{\ss}e 15, D-12489 Berlin}

\pacs{87.14.Gg}{DNA, RNA}
\pacs{02.50.Ey}{Stochastic processes}
\pacs{82.37.-j}{Single molecule kinetics}

\begin{document}

\maketitle

\begin{abstract}
We consider the diffusive motion of a localized knot along a linear polymer
chain. In particular, we derive the mean diffusion time of the knot
before it escapes from the chain once it gets close to one of the chain ends.
Self-reptation of the entire chain between \emph{either\/} end and the knot
position, during which the knot is provided with free volume, leads to an
$L^3$ scaling of diffusion time; for sufficiently long chains, subdiffusion will
enhance this time even more. Conversely, we propose local ``breathing'', i.e.,
local conformational rearrangement inside the knot region (KR) and its
immediate neighbourhood, as additional mechanism. The contribution of
KR-breathing to the diffusion time scales only quadratically, $\sim L^2$,
speeding up the knot escape considerably and guaranteeing finite knot
mobility even for very long chains.
\end{abstract}

\date{\today}

\section{Introduction}

\begin{figure}
\twoimages[height=6.6cm]{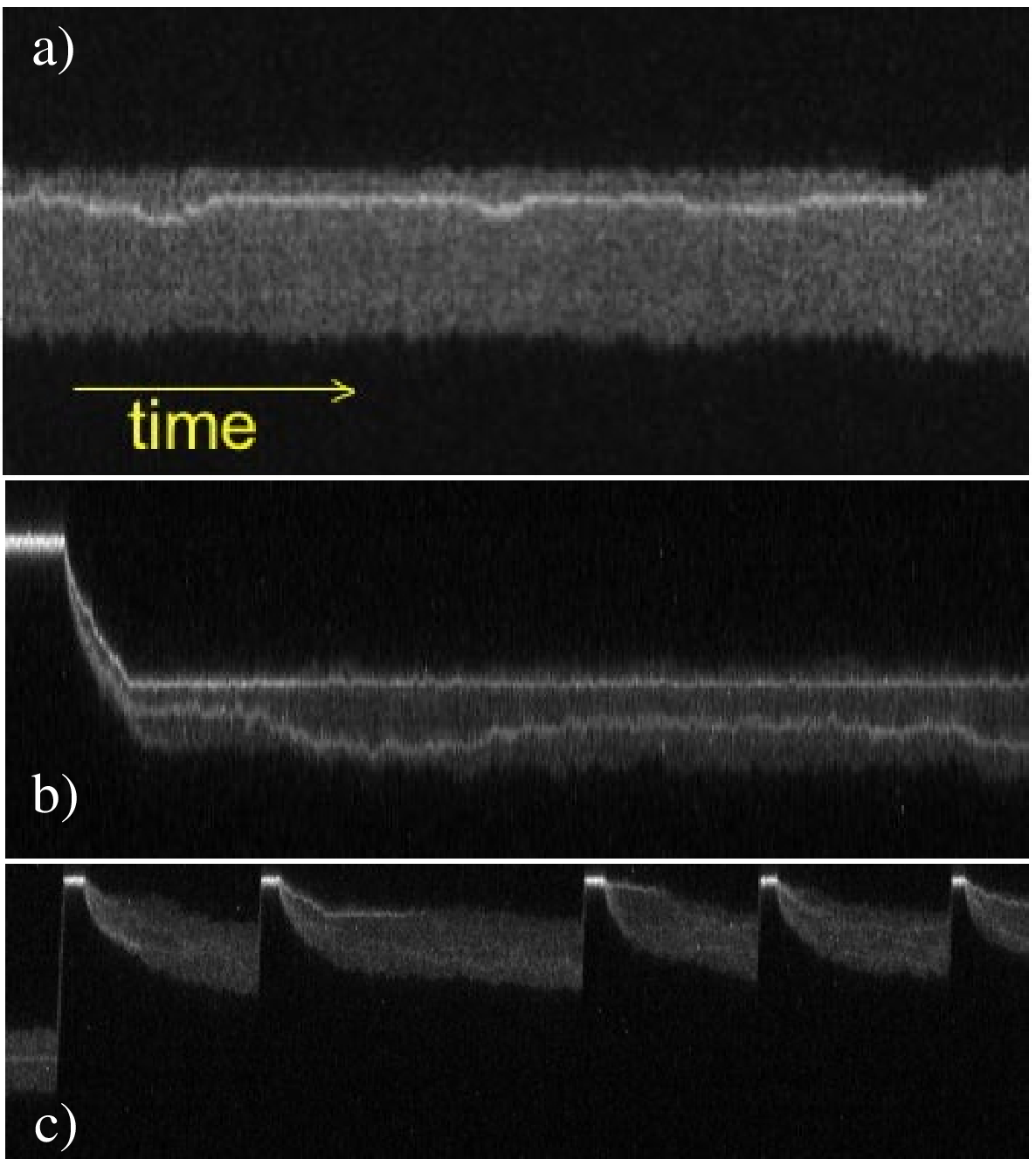}{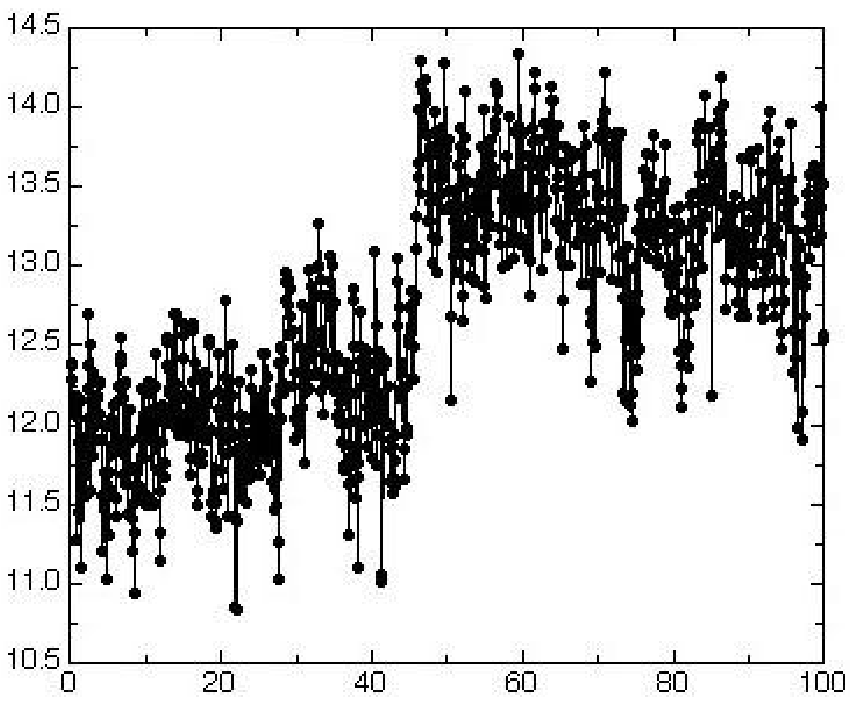}
\caption{Phage $\lambda$ DNA of contour length $L\approx 16.5\mu$m in a
60$\times$80nm channel. Left: Three fluorescence time-traces of the DNA.
a) The feature is almost immobile before escape; b) Two features created
by a forced collision; c) repeated collisions producing new features, that
escape prior to a new collision. Right: Evolution ([$\mu$m] versus [sec])
of the end-to-end length corresponding to the trace in a).
The jump ($\Delta N\approx 2.3\mu$m, or 6.8kbp) at around 42 sec corresponds
to the escape of the feature \cite{walter}.}
\label{knotfig}
\end{figure}

Double-stranded DNA under standard salt conditions has a persistence length
$\ell_p$ of \emph{circa\/} 50nm. Nanochannels can presently be manufactured
with widths that are comparable to, or even smaller than, $\ell_p$. In such
channels, that is, linear DNA is stretched beyond the blob regime, assuming
undulations in the bending-affected Odijk regime \cite{odijk}. This makes it
possible to detect and separate individual DNA molecules and characterise,
for instance, their length \cite{tegenfeldt,reisner}; it even appears feasible
in the future to use nanochannels for sequence determination \cite{austin}.
While being pushed through the nanochannel by an electric field, the DNA chain
occasionally collides with local channel defects, bottlenecks created during
the manufacturing process. Such collisions often cause a superstructure in
the chain, visible in the experiment using stained DNA as a spot of increased
fluorescence and concurrent reduction of the end-to-end length of the chain
(figure \ref{knotfig}). On switching off the electric field, after a period of
immobilisation, these features eventually move toward one of the chain ends
and escape \cite{walter}. We argue that these intermittent features likely
correspond to knots threaded into the DNA on collision.\footnote{Repeated
collision of the same chain with the defect almost always causes a new
superstructure (figure \ref{knotfig}).} In what follows, we propose a
mechanism for the diffusive escape of the knot, obtaining a shorter than
previously assumed knot diffusion time.

Knots also occur \emph{in vivo}. For instance, in bacteria cells DNA knots are
created physiologically \cite{deibler}; they are implicated to have an active
role in gene regulation, separating topologically different parts of the
genome \cite{gene}. Packaging of mutant DNA into virus capsids turns out to be
a highly efficient manner to produce DNA knots in vitro \cite{arsuaga}. After
dissociation of the capsid, the knot needs to be locked by cyclisation of the
formerly linear DNA, before it escapes at either chain end. In both cases it
is of interest to characterise better the knot motion along the DNA chain.
Nanochannel techniques promise to add novel quantitative information to the
understanding of knot motion in polymers.

Knot diffusion in stretched, single $\lambda$ DNA was studied experimentally
by Bao \etal~\cite{quake}. In their analysis they argue that the diffusion
time of the knot from an initial position inside the chain to one of the ends
should scale like $L^2/D$. Assuming a diffusion coefficient $D\simeq L^{-1}$,
the diffusion time $\tau_D$ of the knot scales like $\tau_D\simeq L^3$. This
scenario implies that the entire chain self-reptates to create the motion of
the knot. We here argue that (i) this self-reptation is asymmetric in an open
knot leading to increase mobility toward the chain ends, and (ii) suggest an
additional mechanism, namely, breathing of the knot region (KR)\footnote{In
the case of a localized knot, the volume that contains all non-trivial
topological entanglements \cite{slili2d}.}. KR-breathing involves exchange of
stored length \emph{locally\/} with the vicinal chain. This mechanism is
expected to dominate in long chains, significantly increasing the knot
diffusivity with quadratic scaling of the diffusion time, $\tau_D\simeq L^2$.
Even in long chains, KR-breathing causes Brownian diffusion.

Let us first discuss the nature of the observed feature. A priori, one might
suspect that the feature is a fold\footnote{Similar to folds created when
trying to push an electric cable in a larger cable canal.} in the DNA of the
shape of a Z or a \rotatebox{90}{$\varphi$}, or a knot. Both simple knot and
fold structures carry a bending energy amounting to some 3$k_BT$.\footnote{A
trefoil knot in a linear chain as well as the Z or \rotatebox{90}{$\varphi$}
folds correspond to the bending energy of a full circle with a radius
equivalent to the channel radius, approx.~50nm. Using $E\approx k_BT\ell_p\pi
/R$ for a radius $R$, we find above value \cite{marko}. Higher order knots,
that are likely to be created, carry higher bending energy \cite{ideal}.}
This is a fairly small
energy such that thermal fluctuations would be sufficient to remove the folds.
Moreover, the folds are not self-confining so that they should fluctuate
significantly in size; on the images, however, the fluorescence
spot is strongly localized at all times. In fact, it is known that in
nano-confined DNA knotting occurs almost with probability one \cite{arsuaga}.
We are therefore confident that the observed features correspond to a knotted
state of the DNA chain. While the knotting studies in nano-confinement suggest
that the created knots are more complex, a defect size of $\approx 2\mu$m
indeed suggests a knot with larger number of essential crossings
\cite{quake,vologodskii}.

\section{Immobilisation period}

Once formed while being smashed against the channel defect by the external
electric field, the created knot is relatively tight such that, initially,
it is immobile. Only when, driven by thermal fluctuations, sufficient
contour length is introduced into the KR, it becomes flexible enough to
start its diffusion toward a chain end. We pursue a simple energetic
argument to quantify the knot immobilisation. In the remainder we assume
that chain relaxation is fast compared to knot immobilisation time.

After creation under high confinement the knot is not molecularly tight in
the picture of de Gennes \cite{degennes_knot}. However, it is small enough
such that the self-confined volume of the KR is of a diameter comparable to
or smaller than $\ell_p$. At several contact points where the topology of
the knot prevents expansion the bending stiffness causes that the respective
chains lock each other, comparable to static friction effects. Fluctuations
now need to simultaneously loosen a number of these locking contacts in
order to allow free length to enter the KR. The energy barrier for releasing
a contact is of the order of the work necessary to displace one part of the
chain from the local potential minimum. To induce a displacement $\delta h$,
the work $\delta E=F_{\perp}\delta h$ needs to be performed, where $F_{\perp}$
is the force acting on the contact area. To determine how $F_{\perp}$ scales
with the size of the KR, $R_{\Delta}$, we note that the typical bending energy
of a DNA segment of length $\ell$ and curvature radius $R_{\Delta}$ is $E_b
\simeq\pi\kappa\ell/(2R_{\Delta}^2)$, with elastic constant $\kappa=k_BT\ell_p$
\cite{marko}. As the KR is not molecularly tight and all segments of the KR
tend to assume minimum curvature, the curvature radius is of the order of
$R_{\Delta}$. A small normal in-plane displacement $\delta h$ of a stretch of
DNA around the contact point changes the curvature radius approximately by
$-\delta h$ (bending reduces $R_{\Delta}$ and there are no other parameters
of the dimension of length in this purely geometric argument), so that
$F_{\perp}\simeq dE_b/d(\delta h)\simeq-dE_b/dR_{\Delta}\sim
R_{\Delta}^{-3}$. Thus, the energy necessary to unlock $n$ contacts is
estimated by $\delta E_n\simeq n\pi(\delta h)\kappa\ell_pR_{\Delta}^{-3}$ if
we assume $\ell$ to be of the order of the Kuhn length $\xi_K$. The sharp
$R_{\Delta}^{-3}$-dependence of the potential barrier for loosening the KR
suggests that the activation energy becomes unimportant after crossing the
first few barriers, and further untightening is controlled by a different
mechanism.

The fast tightening of a knot into a locked state by an external force
$\zeta$ requires the crossing of practically the same potential barrier.
If the associated displacement in direction of the force is $b$, the work
necessary is $\delta E_{\zeta}\simeq b\zeta$.\footnote{This relation is
analogous to de Gennes' result for a molecularly tight knot
\cite{degennes_knot}.} The time scale for knot immobilization according
to our barrier crossing argument is therefore (compare \cite{degennes_knot})
\begin{equation}
\theta\simeq\tau\exp\left[\delta E_n/k_BT\right],
\label{eq: degennes}
\end{equation}
where $\tau$ is a relaxation time, not necessarily of the entire chain. Due to
the sharp $R_{\Delta}^{-3}$ dependence of the barrier $\delta E_n$, it rapidly
decays when the size of the KR grows. Therefore the immobilization time of a
knot in an initially tight configuration is large (and proportional to the
exponential of the initial tightening force $\zeta$), but the thermally
activated untightening process starting after overcoming the initial
potential barrier is extremely fast and avalanche-like. After a few thermally
activated steps the knot takes its maximally released configuration, with the
size restricted only by the external geometric constrains (i.e., channel or
tube diameter). The next stage of the process is therefore not
activation-dominated, but corresponds to a diffusion of the stored length
into the KR. In a control experiment on a linear chain, it should
be possible to measure $\tau$ independently. Therefore one could obtain the
value of the activation energy $\delta E_n$ as function of the electric
field smashing the knot against the channel defect, or, in the case of
optical tweezers experiments, the mechanical tension $\zeta$. Note that for
some of the specific knots recorded, the immobilisation period lasts for
several tens of seconds.
The number of contact points increases with knot complexity, while the
curvature radii become somewhat smaller, such that more complex knots
are expected to have a drastically enhanced immobilization time.

\section{Confined knot}

Once unlocked by the provision of sufficient stored length, in absence of
a strong line tension the knot remains in the loose mode until it escapes
from the chain. By exchanging stored length between the KR and the rest of
the chain, the knot performs a random walk along the polymer. As we will see,
the ensuing diffusion process is biased. We distinguish two potential
diffusion mechanisms, both providing the fluctuations in additional length,
leading to an inchworming of the KR:
(i) The first diffusive mode is characterised by a strongly
disbalanced behaviour, in which length is provided to the KR by motion of the
entire chain (from KR to the respective end) to either side of the knot. This
`longitudinal' mode is similar to the self-reptation picture of \cite{quake},
however, our assumption takes the asymmetry into consideration, that the
self-reptative diffusivity should increase when the knot gets closer to
one of the chain ends. The efficiency of self-reptation should become reduced
under tension (straightening
of the chain). (ii) The second mode is an autonomous motion of the KR, due to
the fact that it is not molecularly tight: local length fluctuations in
exchange with the knot's immediate vicinity slightly change its size, allowing
random directional changes. We refer to this `transversal' mode consisting of
small variations in the knot size as KR-breathing, causing a sliding motion
of the KR along the chain. Once the knot arrives at one chain end, it is
released, corresponding to a first exit problem with absorbing boundaries at
$x_L=0$ and $x_R=0$.

Both modes of knot release imply coherent motions of considerable parts of the
chain. We assume that such concerted motion of a part of the chain of length
$l$ invokes the Rouse-like mobility $\mu(l)\propto l^{-1}$, so that the
corresponding diffusion coefficient is $D(l)=\mu(l)k_B T \simeq k_BT/l$.
Thus, if $\Delta \ll L$, KR-breathing typically involves the motion
of a much shorter segment (of length $\Delta$) than self-reptation, unless the
knot travels close to the end of the chain, when the second mode becomes
comparable or even more relevant.\footnote{In particular, for an unconfined
knot as discussed below.} Let us now discuss the two types of motion in more
detail. KR-breathing is characterised by a diffusivity
\begin{equation}
D_{\mathrm{KRB}}\simeq \mu_{\mathrm{KRB}}(\mathcal{K})k_BT\simeq\alpha k_BT/
\Delta(\mathcal{K})
\end{equation}
involving the motion of a length $\approx\Delta$ of the KR and a small vicinal
DNA. The mobility $\mu_{\mathrm{KRB}}\simeq1/\Delta$ will strongly depend on the
knot type $\mathcal{K}$ through its size $\Delta(\mathcal{K})$ \cite{quake}.
In particular, $D_{\mathrm{KRB}}$ is independent of the overall chain length,
$L$. In scaling terms, the diffusion time of the knot on the chain is given by
$\tau_{\mathrm{KRB}}\simeq L^2/D\simeq\Delta L^2$, growing slower than the
cubic scaling of the reptation time for chain dynamics in a reptation tube
\cite{degennes,doi}. To obtain the dependence of the knot diffusion time on its
initial location $x_L$ on the chain, we determine the mean first passage
time $\mathcal{T}(x_L)$ from the diffusion equation
$\partial P(x,t)/\partial t=(\partial/\partial x)D(x)(\partial/\partial x)
P(x,t)$, with absorbing boundary conditions at both ends. Here, $P(x,t)$ is
the probability density to find the defect at position $x$. Following
Gardiner \cite{gardiner}, $\mathcal{T}(x_L)$ is given by
\begin{equation}
\label{mfpt}
\mathcal{T}(x_L)=\int_0^{x_L}\frac{L_{\mathrm{red}}/2-x'}{D(x')}dx',
\end{equation}
where $L_{\mathrm{red}}=L-\Delta$ is the reduced length of the two linear
segments of the chain. With the expression for $D_{\mathrm{KRB}}$, the
diffusion time for KR-breathing yields:
\begin{equation}
\mathcal{T}_{\mathrm{KRB}}
(x_L)=\frac{\Delta L_{\mathrm{red}}^2}{2\alpha k_BT}\left[\left(x_L/L_{\mathrm{
red}}\right)-\left(x_L/L_{\mathrm{red}}\right)^2\right].
\label{diff1}
\end{equation}

\begin{figure}
\includegraphics[scale=0.554]{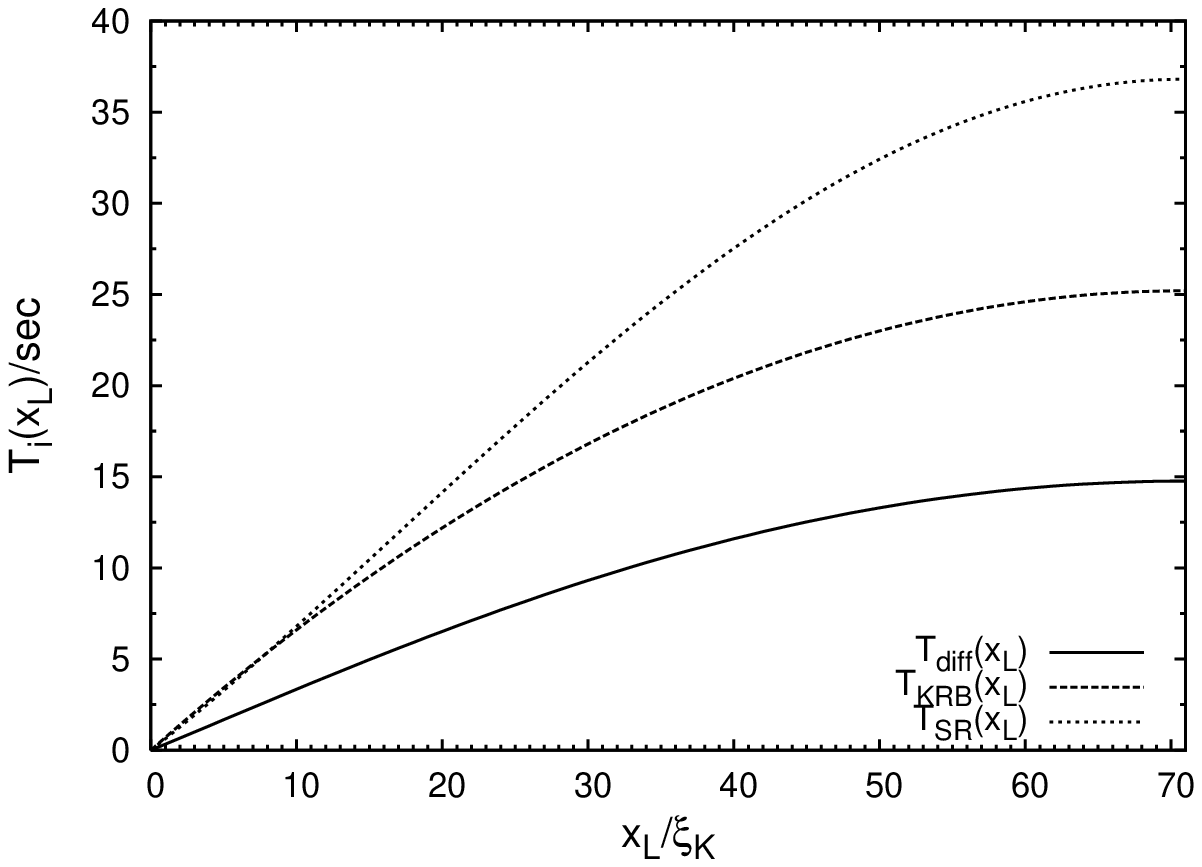}
\includegraphics[scale=0.554]{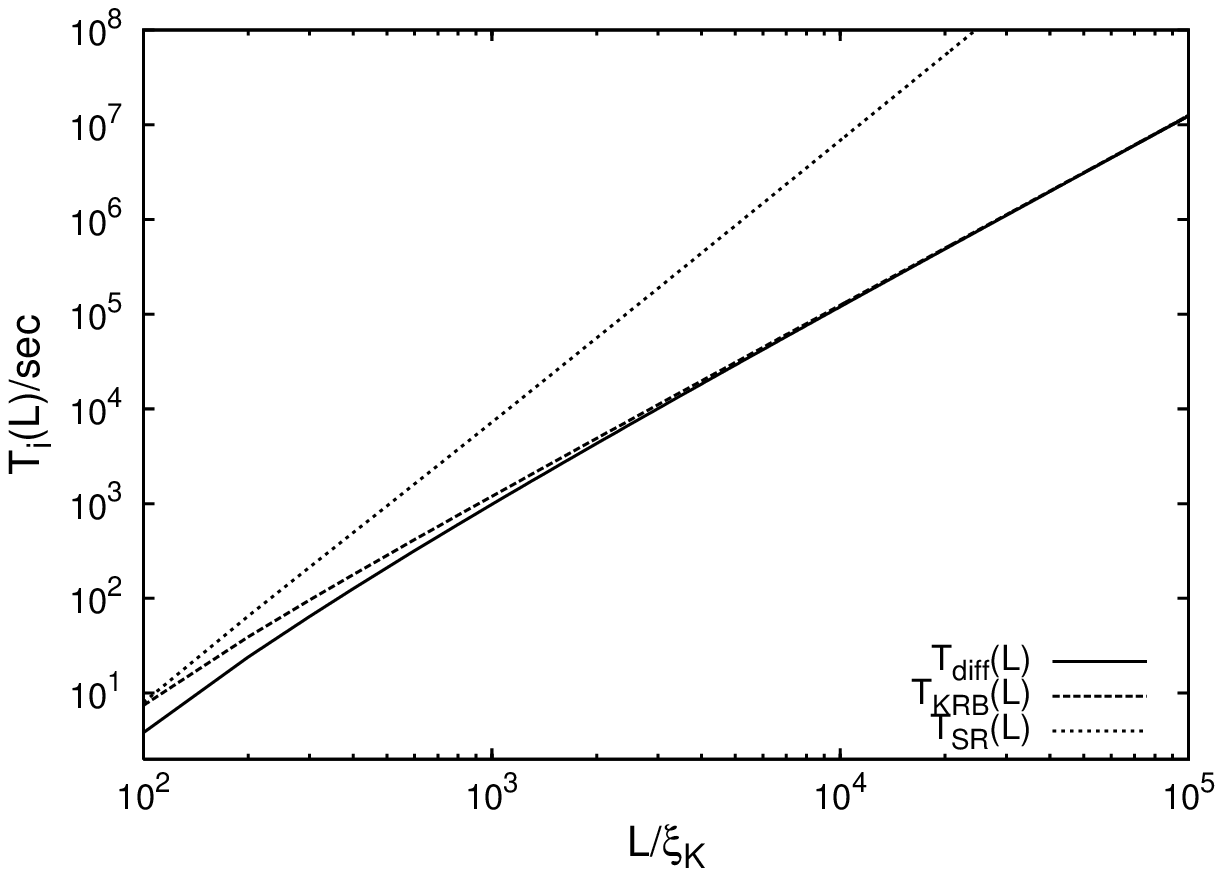}
\caption{Mean first passage time $\mathcal{T}(x_L)$ versus initial position
$x_L$ of the knot for fixed chain length $L=165\xi_K$ (left), and versus chain
length $L$ for symmetric initial condition $x_L=L_{\mathrm{red}}/2=71\xi_K$
(right). With $\Delta=23\xi_K$ and $\tau_K=10^{-2}$sec.}
\label{mfpt_fig}
\end{figure}

In the self-reptation picture, the second diffusion mode, KR motion requires
rearrangement of either the left or the right chain respective to the knot
location plus the KR itself. Thus,
\begin{equation}
D_{\mathrm{SR}}\simeq\frac{\alpha k_BT}{x_L+\Delta}+\frac{\alpha k_BT}{x_R+
\Delta}=\frac{\alpha k_BT(L+\Delta)}{(x_L+\Delta)(L-x_L)},
\end{equation}
that quickly increases toward either chain end. From expression (\ref{mfpt}),
it follows
\begin{eqnarray}
\nonumber
\mathcal{T}_{\mathrm{SR}}(x_L)=\frac{1}{12\alpha k_BT(L+\Delta)}&\Big[
x_L\Big\{x_L\big(3LL_{\mathrm{red}}-4Lx_L-2L_{\mathrm{red}}x_L+3x_L^3\big)\\
&+\Delta\big(6L\big[L_{\mathrm{red}}-x_L\big]
+x_L\big[4x_L-3L_{\mathrm{red}}\big]\big)\Big\}\Big]
\label{diff2}
\end{eqnarray}
plotted in figure \ref{mfpt_fig}. If $\Delta\ll L$, we have $L\approx
L_{\mathrm{red}}$, and we find the simplified expression
\begin{eqnarray}
\mathcal{T}_{\mathrm{SR}}(x_L)\approx\frac{L^3}{4\alpha k_BT}
\left[\left(x_L/L\right)^2-2\left(x_L/L\right)^3+\left(x_L/L\right)^4\right].
\end{eqnarray}
Self-reptation, that is, leads to the $L^3$ scaling of $\mathcal{T}_{
\mathrm{SR}}(x_L)$
in analogy to regular reptation \cite{degennes,doi}.

Combining both effects to the total diffusivity $D_{\mathrm{diff}}(x_L)=D_{\mathrm{KRB}}(
x_L)+D_{\mathrm{SR}}(x_L)$, we cannot obtain an analytic result for the knot
diffusion time and resort to numerical evaluation.\footnote{Noting that in
reference \cite{quake}, for a
slightly pulled chain the diffusivity $D_{\mathrm{diff}}\approx1.2\mu\mathrm{
m}^2/$sec was measured.} Rewriting $D_{\mathrm{diff}}=\xi_K^2/\tau_K$ in terms
of the Kuhn length as assumed unit for stored length and therefore the
fundamental random walk step size, for $D_{\mathrm{diff}}\sim1\mu\mathrm{m}^2
/\mathrm{sec}$, we find the associated step time $\tau_K\sim 10^{-2}$sec.
From above relations we obtain the graphs in figure \ref{mfpt_fig}. 
For initial positions of the KR close to a chain end, the mean first passage
times for both diffusion mechanisms becomes identical, as it should. For a
chain of the size of the $\lambda$ DNA ($L=165\xi_K$), the individual modes
are both relevant. The scaling of $\mathcal{T}_{\mathrm{diff}}$ with the chain
length for the symmetric initial condition $x_L=L_{\mathrm{red}}/2$ is shown
in figure \ref{mfpt_fig}, demonstrating that for longer chain lengths,
KR-breathing dominates. The scaling dependence of the overall knot diffusion
time provides a direct way to test our model experimentally.
It will be interesting to monitor the time series of knot defects in a large
ensemble of realizations to obtain statistically relevant results for the two
contributions $\theta$ and $\mathcal{T}_{\mathrm{diff}}$, i.e., immobilisation
and diffusion times.

\section{Flexible chain}

For flexible chains with persistence lengths much smaller than the channel
diameter, the KR is allowed to expand. The size of (at least, simpler)
knots in
absence of tension, or under moderate tension scales like a power-law of the
chain length, $\ell_k\simeq a^{1-c}L^{c}$, with $c\approx 0.7$
\cite{virnau,stella}. The modus operandi of knot escape from the chain
can now follow two different tracks: (i) First, the knot can diffuse out
while being localized at all times, corresponding to the scenario developed
above. As $\xi_K$ is much smaller and therefore $L$ much larger in units of
$\xi_K$, KR-breathing dominates, and the first passage time scales like
$\mathcal{T}_{\mathrm{diff}}\simeq L^2/D_{\mathrm{KRB}}$. Without effective
confinement by the channel walls, the KR grows with $L$, entering the
diffusivity $D_{\mathrm{KBR}}=D_{\mathrm{KBR}}(L)$, i.e., the diffusive
component with fractal scaling $\mathcal{T}_{\mathrm{KRB}}\simeq L^{2+\alpha}$
yields.
For knots dominated by finite size effects (or complex knots that might tend
to delocalisation) \cite{quake1}, or under dense or $\theta$ conditions
\cite{hanke}, the KR scales like the overall chain length, and we
would obtain the reptation scaling $\mathcal{T}_{\mathrm{KRB}}\sim\mathcal{T}_
{\mathrm{diff}}\simeq L^3$.
Conversely, when confined to
two dimensions by adhesion to a surface or by sandwiching between two 
parallel walls, the knot localises tightly and $\alpha=0$ \cite{slili2d}.
(ii) Second, in absence of external tension the knot can be undone by
swelling of the KR until it is of the size $L$ of the chain, and the chain
ends are inside the KR. This mode is expected to scale like the all-chain
Rouse time $\simeq L^3$.

\section{Discussion}

The possibility to devise nanochannels of a width comparable to the
persistence length of polymers, and to monitor the end-to-end distance
of a single polymer is a fine example how new experimental techniques
can probe mechanically domains of fundamental behaviour of single
polymers. We presented a simple scaling model of both the immobilisation and
the diffusion of knots in DNA confined in nanochannels, identifying two
different reptation modes. While self-reptation has been considered earlier,
we propose a modified self-reptation picture that is, in fact, asymmetric. We
also suggest a new mode, namely, KR-breathing, that is expected to dominate
the knot diffusion in longer chains. By varying chain length $L$ and the
channel diameter-to-persistence length ratio, we expect future experiments
to provide new insight in the reptation of single chains that self-confine
themselves.

It appears obvious that when the knot gets close to one of the chain ends,
the close end can easily reptate into the KR and subsequently release the
knot, without the far end being involved. In this picture, it is not
necessary to have full chain equilibration on time scales slower than
the diffusion time. The connected asymmetry of the diffusion coefficient
with respect to the knot position along the chain is indeed corroborated
by recent simulations results of a trefoil knot diffusing on a linear chain
under moderate tension \cite{zimmermann}. The existence of two diffusion
modes makes sure that the knot mobility is finite even for very long chains,
$L\to\infty$. In other pictures assuming relaxation of the whole chain as
relevant diffusion time scale, in a very long polymer the knot would become
effectively immobile.

We note that small external stretching forces should not affect the knot
diffusion modes. Moderate forces are expected to reduce self-reptation; as
long as the blob size remains larger than the KR, KR-breathing should not
be disturbed. For high forces, eventually the system will be immobilised
according to the above picture. We also note that we assumed that
the diffusion of the knot modes is Brownian. While this is always true for
the KR-breathing mode, for long chains self-reptation should become
subdiffusive according to arguments brought forth in the discussion of
nanopore translocation \cite{kantor,mekla}. In that case, KR-breathing becomes
even more dominant. It would be very interesting to directly observe
subdiffuion in experiments.

Lastly, from the experimental side, it appears that the de Gennes hypothesis regarding knot immobilization time (Eq. \ref{eq: degennes}) could be tested by varying the field strength used to compress the DNA, as the tightening force should be related to the compression field.  The de Gennes hypothesis would suggest that the knot immobilization time is a strongly increasing function of the compression field.  Also, it would be interesting to study how the generated knot spectrum depends on compression field strength and duration.  This could easily be accomplished by compressing DNA at varying field strengths.

\acknowledgments

RM acknowledges the Natural Sciences and Engineering Research Council (NSERC)
of Canada and the Canada Research Chair program for support. For the
experimental work, which was carried out at Prof. Austin's group at Princeton,
we would like to acknowledge support from NIH (HG01506) and NSF Nanobiology
Technology Center (BSCECS9876771).

\end{document}